\documentclass[multphys,vecphys]{svmult}

\usepackage{makeidx}         
\usepackage{graphicx}        
\usepackage{multicol}        
\usepackage[bottom]{footmisc}

\makeindex             

\begin{document}

\title*{3D Spectroscopy of Herbig-Haro objects}
\titlerunning{3D HHs Spectroscopy} 

\author{R. L\'opez\inst{1}\and
S.F. S\'anchez\inst{2}\and
B. Garc\'\i a-Lorenzo\inst{3}\and
R. Estalella\inst{1}\and
G. G\'omez\inst{3}\and \\
A. Riera\inst{4}\and
K. Exter\inst{3}}
\authorrunning{R. L\'opez et al.} 

\institute{Dept. d'Astronom\'\i a i Meteorolog\'\i a. U. Barcelona.
\texttt{rosario,robert@am.ub.es}
\and Centro Astron\'omico Hispano-Alem\'an de Calar Alto.  \texttt{sanchez@caha.es}
\and Instituto de Astrof\'\i sica de Canarias \texttt{bgarcia,ggv,katrina@ll.iac.es}
\and Dept. F\'\i sica i Enginyer\'\i a Nuclear (EU Vilanova i la Geltr\'u , UPC)
\texttt{angels.riera@upc.edu}}
\maketitle

\section{Introduction and Observations}

HH 110 (in L1267, $d=460$ pc \cite{lopez:Rei91}) and HH 262 (in L1551, $d=140$ pc
\cite{lopez:Graham90}) 
are two Herbig-Haro (HH) objects that
share a peculiar, rather chaotic morphology. In addition, no stellar source powering these jets has been
detected at optical or radio wavelengths. Both, previous observations
\cite{lopez:Lopez98} \cite{lopez:Lopez05}
and models \cite{lopez:Raga02}, suggest that the
jet emission reveals an early stage of the interaction between a supersonic outflow and the dense
outflow environment. These HHs are thus suitable to search for observational signatures of
supersonic outflow/dense environment interaction.

We mapped these HHs with the Integral Field Instrument PMAS (Postdam Multi-Aperture
Spectrophotometer) at the 3.5m CAHA telescope, under the PPAK configuration (331 science fibers,
of 2\rlap.''7 each, in an hexagonal grid of $\sim$ 72'' of diameter). We used the J1200 grating
(spectral resolution $\sim$ 15 km~s$^{-1}$ for H$\alpha$; wavelength range $\sim$ 6500-7000 \AA\,
that includes the emissions from the characteristic red HHs lines: H$\alpha$, and the [NII] 
and [SII] doublets).
Mosaics from several overlapping pointings (4 for HH 110 and 8 for HH 262) were made in order to cover
the whole area of the emission of the HHs.

\section{Results}
From the PMAS data, we obtained maps of the morphology (monochromatic flux) 
and kinematics (radial velocity field and
velocity dispersion) for the H$\alpha$, [NII] and [SII] line emissions of these two HHs. 
In addition, we computed line-ratio maps in order to explore the 2D structure of the electron
density and excitation conditions, looking for the behaviour of the excitation and density spatial
structures as a function of the velocity field. We show in the next section some of the  maps obtained
as an exemple of the PMAS science output.
\subsection{HH 110}

The morphology of the emission is very similar for all the lines mapped, although some differences among
the three lines can
be appreciated. In particular, the H$\alpha$ emission appears sligthly more extended across the jet
beam than the [NII] and [SII] emissions (a low-emission component surrounding the knots is only detected 
in the H$\alpha$ line). 

The radial velocities derived from all the lines mapped appear blueshifted (in the rest frame 
of the cloud) and show a similar behaviour in all the lines along the jet 
axis: first, $V_R$ becomes  redder as we move from the northern edge of the
jet (knot A, with  $V_R$ $\simeq$ $-$30 km~s$^{-1}$ in H$\alpha$) to the south, up to a 
distance of $\sim$ 40'' from knot A, 
where the reddest $V_R$ values are found ($V_R$ $\simeq$ $-$15 km~s$^{-1}$ in H$\alpha$); it changes
further out towards more blueshifted values; the
highest blueshifted values of $V_R$ ($\simeq$ $-$35 km~s$^{-1}$ in H$\alpha$) are found at $\sim$ 100'' 
from knot A.

In general, electron densities, as derived from the sulphur line ratio, decrease along the jet axis 
with 
distance from knot A. However, the highest values ($n_e$ $\simeq$ 700 cm$^{-3}$) are found at $\sim$ 40''
from knot A, around the positions  where $V_R$ reaches the  
reddest values for all the mapped lines.

\begin{figure}
\centering

\includegraphics[height=8cm]{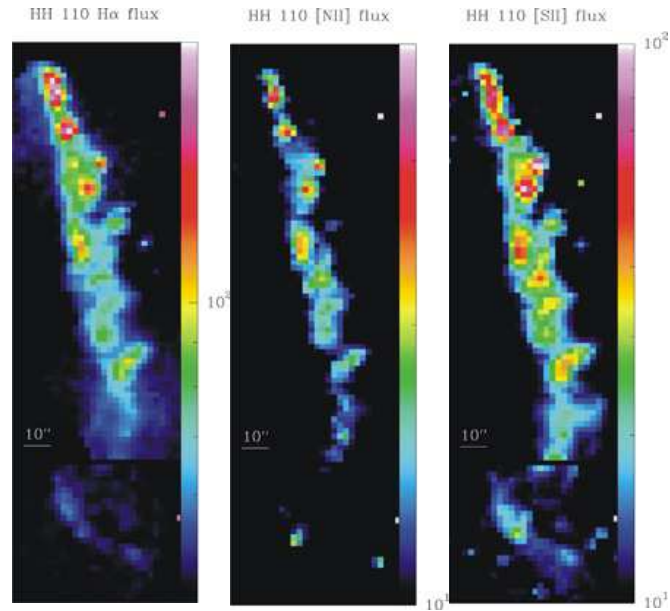}

\caption{HH 110: Monochromatic maps of the integrated flux in the H$\alpha$, [NII] and [SII] emission
lines. North is up and East is to the left.}
\label{lopez:fig 1}       
\end{figure}

\subsection{HH 262}
The morphology of the H$\alpha$ and [SII] emissions are similar. In contrast, emission from the
[NII] line is only found in the northwestern region of HH 262.

The radial velocities derived from the H$\alpha$ and [SII] lines appear redshifted, in the rest frame of the cloud,
for all the HH 262 knots, and present a rather complex
pattern. As a general trend, $V_R$ increases from the north to the center of HH 262 and decreases from the center
to the south. For the HH 262 knots, 
significant differences between the H$\alpha$ and [SII] $V_R$ values are found, and the velocity
difference changes for each knot. FWHMs of these two lines appear significantly wider (by 20--30
km~s$^{-1}$) towards the center of HH 262 relative to the rest of the knots.

Electron densities derived from the sulphur line ratios are close to the low-density limit for most of
the HH 262 knots. The highest values ($n_e$ $\simeq$ 240 cm$^{-3}$) are found towards the
northwestern HH 262
knots, coinciding with the loci where  emission from the [NII] lines is detected. 
For the rest of the object, electron density  is less than 
100 cm$^{-3}$, decreasing slightly from north to south.

\index{paragraph}

\begin{figure}
\centering

\includegraphics[height=6cm]{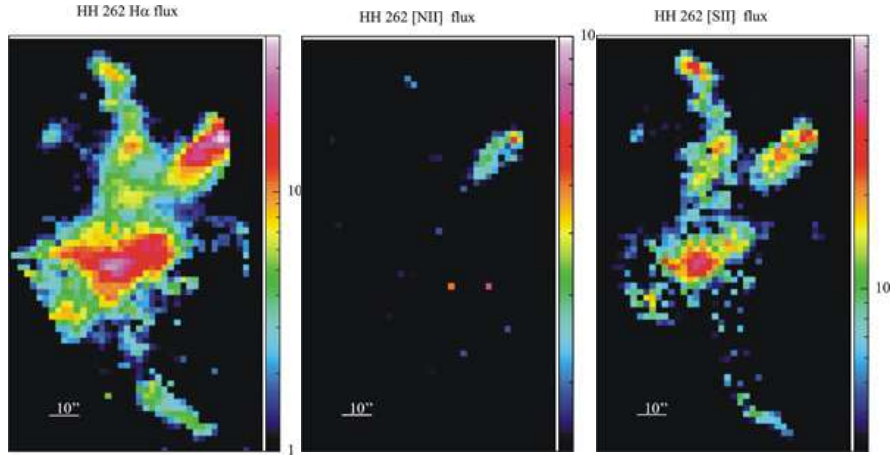}
\caption{HH 262: Same as Fig. 1 but for HH 262}
\label{lopez:fig 2}       
\end{figure}




\printindex
\end{document}